\documentclass[runningheads]{llncs}
\usepackage{graphicx}
\usepackage{amsmath}

\begin{document}
\title{Hybrid Classical-Quantum method for Diabetic Foot Ulcer Classification}
\author{Azadeh Alavi\inst{1,2} \and
Hossein Akhoundi\inst{3}}
\institute{AI Discipline, School of Computing Technologies, RMIT University, VIC 3001, Australia 
\email{azadeh.alavi@rmit.edu.au} \and Bioinformatics lab, Baker Heart and Diabetes Institute, VIC 3002, Australia
\and Security Business Unit, Cisco, Melbourne}

\maketitle


%
%
%
%
%
\begin{abstract}
 Diabetes is a raising problem that affects many people globally. Diabetic patients are at risk of developing foot ulcer that usually leads to limb amputation, causing significant morbidity, and psychological distress. In order to develop a self monitoring mobile application, it is necessary to be able to classify such ulcers into either of the following classes: Infection, Ischaemia, None, or Both. In this work, we compare the performance of a classical transfer-learning-based method, with the performance of a hybrid classical-quantum Classifier on diabetic foot ulcer classification task. As such, we merge the pre-trained Xception network with a multi-class variational classifier. Thus, after modifying and re-training the Xception network, we extract the output of a mid-layer and employ it as deep-features presenters of the given images.  Finally, we use those deep-features to train multi-class variational classifier, where each classifier is implemented on an individual variational circuit. The method is then evaluated on the blind test set DFUC2021. The results proves that our proposed hybrid classical-quantum Classifier leads to considerable improvement compared to solely relying on transfer learning concept through training the modified version of Xception network.

\end{abstract}
\section{Introduction}

Diabetes is a raising universal problem that affects 425 million people which is expected to rise to 629 million people by 2045 \cite{info,DFUC2021}. One in three diabetic patients are likely to develop Diabetic Foot Ulcers (DFU) which is a serious complication of diabetes, and can lead to limb amputation, or even death if it is with infection and ischaemia \cite{info2}. 
However, as diabetic patients can loose sensation in their foot, it is hard for them to identify the development of such ulcer. In an effort to develop a self monitoring technology, in this work we study the classification of such ulcers into following classes : Infection, Ischaemia, None, Both.

\begin{figure}
\includegraphics[width=\textwidth]{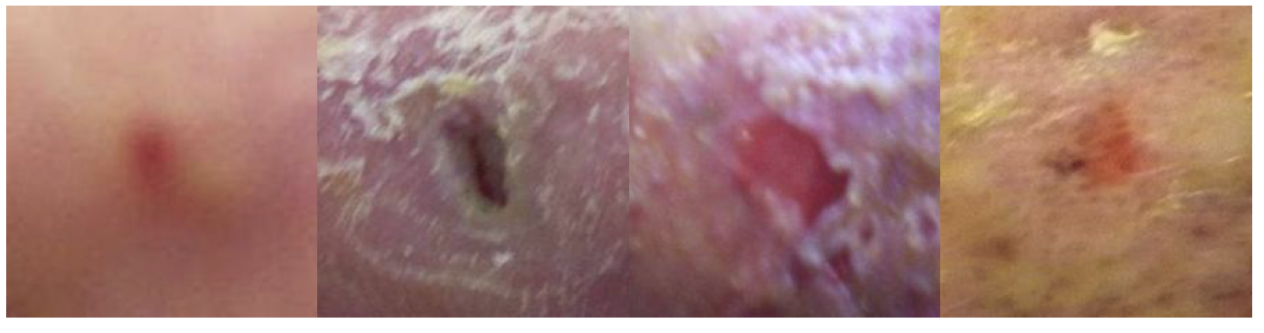}
\caption{An example of early stage DFU \cite{DFUC2021} } \label{fig1}
\end{figure}

Specifically, in this work, we propose a hybrid Classical-Quantum solution for Classification (CQC) of diabetic foot ulcer images (DFU). The proposed classifier is constructed of 2 separate algorithms. First, we employ the concept of transfer learning by adopting Xception Network with pre-train image-net weights \cite{Xception}. Then, we modify and re-train the network to adopt it for DFU classification task. Next, we extract an output of a mid-layer to represent each image. Finally, we employ a multi-class variational classifier \cite{Pref1,Pref2,pytorch,PennyLane}, where each classifier is implemented on an individual variational circuit. To train the classifier, we use the output of the previous stage as image repesentatives. \\

We have evaluated our method on DFU2021 blind test set, and the result shows promising performance. The result proved that the proposed hybrid method achieves considerably higher performance in comparison with the performance of modified Xception net.

\section{Image analysis with Quantum Machine Learning} 

Quantum machine learning employs the principles of quantum theory to process information through using quantum computers. In recent years, we have witnessed the remarkable breakthroughs in quantum technology, as quantum computing has entered the era of Noisy Intermediate-Scale Quantum \cite{conf:refQ1}. As a result, recently the study of Quantum Machine Learning (QML) techniques have became a subject of interest in the research field \cite{conf:refQ2,conf:refQ3}.

Among the applications of QML, quantum image classification has shown promising performance. For example, Llyod et al \cite{conf:refQ4} proposed a method for training the map from classical data to quantum states. Their method demonstrated success in classifying given images into two classes: ant and bee.  In \cite{conf:refQ5}, different quantum parametrized circuits were employed to design number of classification methods, where the classical data were encoded as separable qubits.

\section{Methodology}

The proposed solution is made of two stages: 1) Employing the concept of transfer learning, and consequently modify and re-train Xception network for DFU classification, and 2) Extracting the output of a mid-layer from modified Xception network, and using it as an input to train a multi-class variational classifier. In the following sections, we will explain the steps in more detail.

\subsection{Stage.1: Transfer Learning} 

We use Extreme Inception (Xception) network with imagenet weights, as our base method ~\cite{Xception}, which  is a deep convolutional neural network based architecture. The main characteristic of the Xception network is decoupling the mapping of cross-channels correlations and spatial correlations in the feature maps of convolutional neural networks. In other words,  Xception network is a linear stack of depth-wise separable convolution layers with residual connections. The Xception network architect have following characteristics: 
\begin{itemize}
    \item Contains 36 convolutional layers. 
    \item The 36 convolutional layers are structured into 14 modules.
    \item Except for the first and the last, all the models have linear residual connections around them.
\end{itemize}

\begin{figure}
\includegraphics[width=\textwidth]{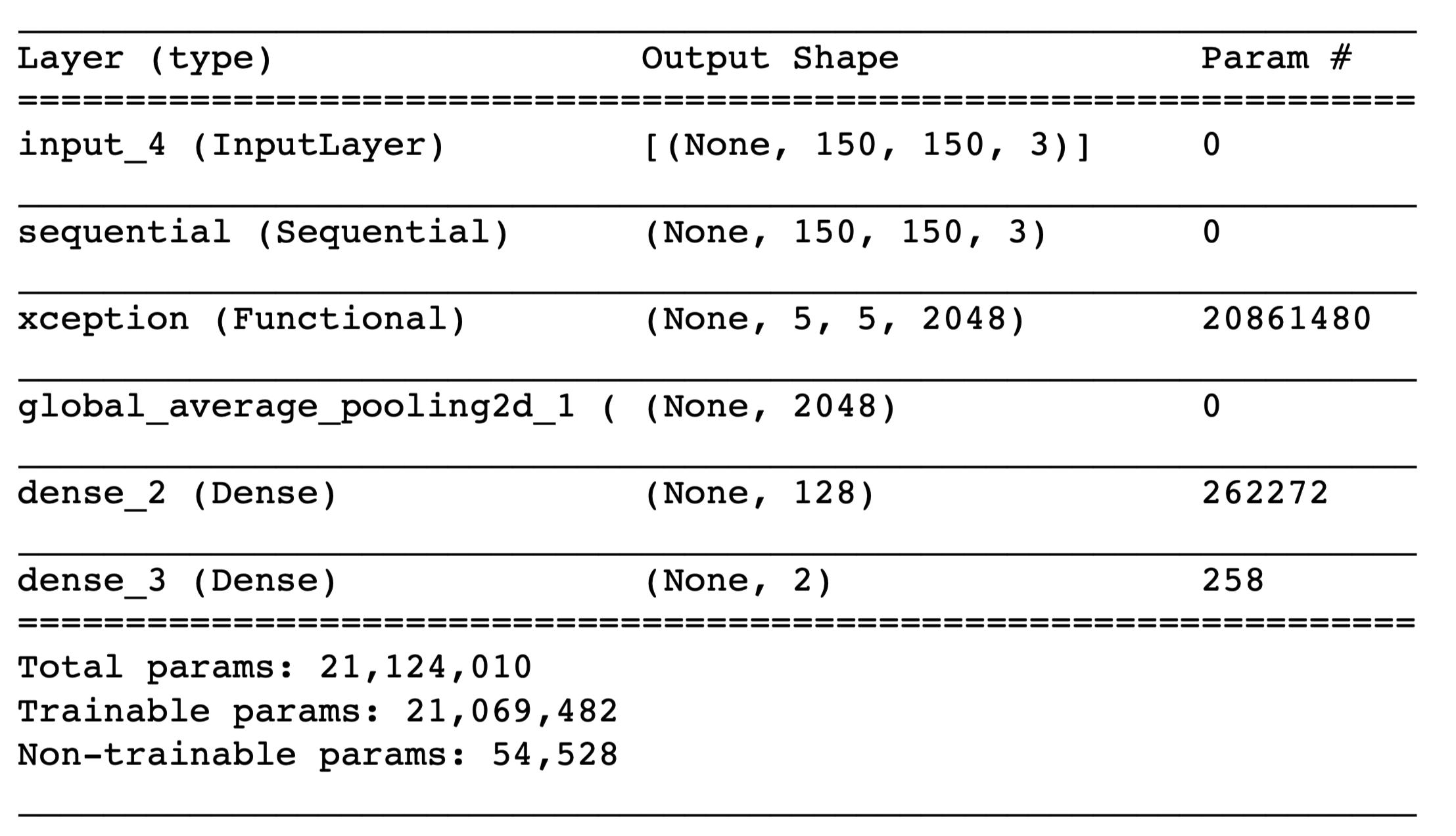}
\caption{The above image provides the detail of the modified Xception network \cite{Xception} } \label{fig1}
\end{figure}

Specifically, in stage.1, we modify Xception network by removing the last layer and adding two fully connected layers of size 128 and 2 to adopt the network for multi-label classification.\\ To address the imbalanced data problem, we increase the penalty weight for Ischaemia class compared to Infection class. 

Then, we train the modified Xception network in two steps: first we freeze the original layers and train the final two layers. Then, we train the entire network with a lower training rate. 
Finally, we extract the output of a mid-layer to represent the descriptive features of each given image.

\subsection{Stage.2: Multi-class Variational Classifier} 

We extract the output of "dense-2 layer" ( Fig. \ref{fig1} ), which is a vector of size 128. Then we use it as a representative feature for training the Multi-class Variational Classifier \footnote{we have employed an elementary circuit for the proof of concept, in the future work we will study more elaborate measurements that can lead to increasing the power of the classifier}. \cite{Pref1,Pref2,pytorch,PennyLane}, explained bellow:

\begin{itemize}
    \item \textbf{Quantum Circuit}
    The first layer is consists of rotation gates ($R_X$, $R_Y$, and $R_Z$) for each qubit, which is then followed by CNOT gates. An example of a rotation gate: 
    \begin{equation*}
    R(\varphi ,\theta ,\omega ) = RZ(\omega) RY(\theta) RZ(\varphi)= \begin{bmatrix}
e^{-i(\varphi +\theta )/2} cos(\theta/2) & e^{-i(\varphi -\theta )/2} sin(\theta/2)\\ 
 e^{-i(\varphi -\theta )/2} sin(\theta/2)& e^{i(\varphi +\theta )/2} cos(\theta/2)
\end{bmatrix}
\end{equation*}
, where $\varphi$, $\theta$, and $\omega$ are the angles for rotation gates.\\
We used 4 QNodes, each representing a one-vs-all classifier as we have 4 classes in total (None, Both, Infection, and Ischameia,). 
    \item \textbf{Loss}
    We employ marginal loss as our loss function; because, our multi-class classifiers is in fact a number of one-vs-all classifier:
\begin{gather*}
L(x,y) = \sum_{i\neq y}^{} max(0, s_i-s_y + \Delta ) \\
~\text{for}~ s_i = c_i(x,\Theta ) ~\text{,where}~ s_i \in [-1,1] 
\end{gather*}
, where $c_i$ is the i th classifier, $x$ is its input, and $s_i$ is its score. 
\end{itemize}
Our model is consist of 6 layers. Finally, we classify the DFU images using the learned model.
\section{Results and Discussion}

To evaluate the proposed method, we have tested its performance on blind test set of Diabetic Foot Ulcers dataset (DFUC2021)\cite{DFUC2021}. We have used 5,955 DFU images for training, 5,734 for blind testing \cite{DFUC2021}. The ground truth labels comprise of four classes: control, infection, ischaemia and both conditions. The results indicate a considerable improvement when using the proposed hybrid CQC compared to solely relying on transfer learning through using the modified version of Xception. Bellow table summarises our finding.

It is important to note that more complex modification of Xception with more data augmentation would result in the better performance for both Xception and CQC. That would be developed and evaluated in future work. Moreover, we would like to emphasise that we have employed an elementary circuit for the proof of concept, in the future work we will study more elaborate measurements that can lead to increasing the power of the classifier.

To generate bellow results, Xception was trained for 10 iterations in the first training step, and 40 for the second stage (with smaller learning rate). 

\begin{table}
\caption{AUC of proposed CQC method validation on blind test set of DFU2021.}\label{tab1}
\begin{tabular}{|l|l|l|l|l|l|}
\hline
Method &  Both-AUC & None-AUC & Infection-AUC & Ischaemia-AUC & Macro-AUC \\
\hline
Modified Xception & 0.6483 & 0.7215 & 0.6438 & {\bfseries 0.7331} & 0.6867 \\
Proposed CQC &  {\bfseries 0.7344} & {\bfseries 0.7282} &  {\bfseries 0.6743} &   0.7163 &  {\bfseries 0.7133 } \\
\hline
\end{tabular}
\end{table}

\begin{table}
\caption{F1-Score of proposed CQC method validation on blind test set of DFU2021.}\label{tab2}
\begin{tabular}{|l|l|l|l|l|l|}
\hline
Method &  Both-F1 & None-F1 & Infection-F1 & Ischaemia-F1 & Macro-AF1 \\
\hline
Modified Xception & 0.3737 & 0.7112 & 0.5503 & {\bfseries 0.5111} & 0.5067 \\
Proposed CQC &  {\bfseries 0.4850} & {\bfseries 0.7153} &  {\bfseries 0.5934} &  0.4411 &  {\bfseries 0.5587} \\
\hline
\end{tabular}
\end{table} 

In Table.1, and Table.2: Both- refers to where the image include both Infection and Ischaemia, None- refers to where image include no Infection or Ischaemia, and Infection- and Ischaemia- refers to where the image include only one condition respectfully. 

The above results indicate that the proposed CQC method results in considerable improvement in performance compared to Xception in general, and in all the area except for classifying Ischaemia. 
The reason is that, while retraining the modified Xception, we have introduced the early stopping option based on monitoring the validation loss. However, training the multi-class variational classifier was solely based on classification accuracy.
We aim to improve our method in the future work to use the validation loss for early stopping, in order to avoid over fitting. 

\section{Conclusion}
Recent breakthrough in the field of quantum computing has led to an increase interest in QML. In this work, we compared the performance of a classical transfer-learning-based method using modified version of Xception network, with the performance of a hybrid classical-quantum Classifier (CQC) on diabetic foot ulcer classification task. As such, we merge the pre-trained Xception network with a multi-class variational classifier.  The method is then evaluated on the blind test set provided by DFU2021 competition. The results proves that our proposed hybrid classical-quantum Classifier leads to considerable improvement compared to solely relying on transfer learning concept through training the modified version of Xception in general, and in all the area except for classifying Ischaemia. 
We have explained that the reason for that is as following: while retraining the modified Xception, we have introduced the early stopping option based on monitoring the validation loss. However, training the multi-class variational classifier was solely based on classification accuracy. We aim to improve our method in the future work to use the validation loss for early stopping, in order to avoid over fitting.
In addition, it is non trivial to note that we have employed an elementary circuit for the proof of concept, in the future work we will study more elaborate measurements that can lead to increasing the power of the classifier.

%
%
%
%
%

\begin{thebibliography}{8}

\bibitem{DFUC2021}
Moi Hoon, Y., Cassidy, B., Pappachan, J., O'Shea, M., Gillespie, D., Neil, R., Analysis Towards Classification of Infection and Ischaemia of Diabetic Foot Ulcers, arXiv preprint arXiv:2104.03068, 2021.

\bibitem{info}
Cho, N., Shaw, J., Karuranga, S., Huang, Y., da J., etc., “Idf diabetes atlas: Global estimates of diabetes prevalence for 2017 and projections for 2045,” Diabetes research and clinical practice, vol. 138, pp. 271–281, (2018).

\bibitem{info2}
Armstrong, D., Boulton, J, Bus, S.,Diabetic foot ulcers and their recurrence, New England Journal of Medicine, vol. 376, no. 24, pp. 2367–2375, 2017.

\bibitem{conf:refQ1}
Yue, R., Xiling, X., Yuanxia, S. Quantum Image Processing: Opportunities and Challenges,Mathematical Problems in Engineering, Machine Learning and its Applications in Image Restoration. vol 2021, Hindawi(2021).

\bibitem{conf:refQ2}
Cappelletti, W., Erbanni, R., Keller, J. Polyadic quantum classifier. arXiv:2007.14044, 2020. 

\bibitem{conf:refQ3}
Cong, I., Choi, S., Lukin, M. Quantum convolutional neural networks. Nature Physics 15, 2019. 

\bibitem{conf:refQ4}
Lloyd, S., Schuld, M., Ijaz, A., Izaac, J., Killoran, N., Quantum embeddings for machine learning, 2020, arxiv.2001.03622.

\bibitem{conf:refQ5}
Grant, E., Benedetti, M., Cao, S., Hallam, A., Lockhart, J., Stojevic, V., Green, A., Severini, S., Hierarchical quantum clas- sifiers. npj Quantum Information 4, arXiv:1804.03680, 2018.

\bibitem{pytorch}
Paszke, A., Gross, S., Massa, F., Lerer, A., Bradbury, J., Chanan, G., … Chintala, S. (2019). PyTorch: An Imperative Style, High-Performance Deep Learning Library. In Advances in Neural Information Processing Systems 32 (pp. 8024–8035). Curran Associates, Inc. Retrieved from http://papers.neurips.cc/paper/9015-pytorch-an-imperative-style-high-performance-deep-learning-library.pdf

\bibitem{Pref2}
  Schuld, M., Bocharov, A., Svore, K., Wiebe, N., Circuit-centric quantum classifiers. Physical Review A, Vol. 101, No. 3, pp. 032308 , ASP(2020).

\bibitem{Pref1}
Edward, F., Hartmut, N., Classification with quantum neural networks on near term processors,arXiv preprint arXiv:1802.06002, 2018

\bibitem{PennyLane}
Ville Bergholm, Josh Izaac, Maria Schuld, Christian Gogolin, M. Sohaib Alam, Shahnawaz Ahmed, Juan Miguel Arrazola, Carsten Blank, Alain Delgado, Soran Jahangiri, Keri McKiernan, Johannes Jakob Meyer, Zeyue Niu, Antal Száva, and Nathan Killoran. PennyLane: Automatic differentiation of hybrid quantum-classical computations. 2018. arXiv:1811.04968
  
\bibitem{ref_QS}
Preskill, J., Quantum computing in the NISQ era and beyond, Quantum, vol. 2, p. 79, 2018.

\bibitem{ref_conf4}
Joachims, T., Transductive inference for text classification using support vector machines, International Conference on Machine Learning (ICML) 1999, Morgan Kaufmann Publishers Inc., pp. 200–209, San Francisco(1999).

\bibitem{ref_article8}
Chen T, Guestrin C, : A scalable tree boosting system, in: Proceedings of the 22Nd ACM SIGKDD International Conference on Knowledge Discovery and Data Mining. ACM, 2016, pp. 785-794.
 

\bibitem{ref_conf1}
Szegedy, C., Vanhoucke, V., Ioffe, S., Shlens, J., Wojna, Z.: Rethinking the inception architecture for computer vision. In: IEEE conference on computer vision and pattern recognition 2016, pp. 2818--2826. IEEE. Nevada (2016).

\bibitem{Xception}
Chollet, F.: Xception: Deep learning with depthwise separable convolutions. In: IEEE conference on computer vision and pattern recognition2016, pp. 1251--1258. IEEE. (2017).

\bibitem{liu2015mlrf}

Feng, L., Zhang, X., Ye, Y., Zhao, Y., Li, Y.: MLRF: multi-label classification through random forest with label-set partition. In: International conference on intelligent computing, pp. 407--418, Springer. (2015)

\end{thebibliography}
%

\end{document}